# RNA Identification and Detection of Nucleic Acids as Aerosols in Air Samples by Means of Photon and Electron Interactions

**John I. Adlish** [1,2]**, Piero Neuhold** [2,] *****, Riccardo Surrente** [2] **and Luca J. Tagliapietra** [2,] *****

[1] Biology Department, Truckee Meadow Community College, Reno, Nevada, NV 89512-3999, USA; john.adlish@babuhawaiifoundation.org
[2] Particle Physics Department, BabuHawaii Foundation, Honolulu, Hawaii 96811, USA.
***** Correspondence: piero@babuhawaiifoundation.org   (P.N.); lucaj@babuhawaiifoundation.org   (L.J.T.)

**Abstract:** This study presents a methodology to reveal traces of viral particles, as aerosol with known chemical and molecular structure, in a sample by means of photon and electron interactions. The method is based on Monte Carlo simulations and on the analysis of photon-electron fluxes-spectra through energy channels counts as a function of different aerosol viral concentrations in the air sample and looking at the peculiar photon/electron interactions with the potential abnormal atomic hydrogen (H), oxygen (O), carbon (C), and phosphorus (P) compositions present in the air sample as a function of living and nonliving matter with PO$_4$ group RNA/DNA strands in a cluster configuration.

**Keywords:** viral particles; radiobiology; RNA detection; microbiology; matter identification; particle physics; physical modelling; particle accelerators in environmental studies

## 1. Introduction

Viruses are intracellular parasites composed of a nucleic acid surrounded by a protein coat, the capsid. Some viruses contain a lipid envelope, derived from the host, surrounding the capsid. The nucleic acid found in viruses can consist of either RNA or DNA. The Coronaviridae, for example contain a single molecule of RNA consisting of about 30 Kilobases in length where Kb stands for unit of measurement of **DNA** or RNA length used in genetics [1, 2-6]. RNA is composed of nucleotides, each containing a sugar (deoxyribose), a Nitrogen containing Base (Adenine, Uracil, Guanine, and Cytosine), and a phosphate group PO$_4$. Members of the family Coronoviridae measure 80-160 nm in diameter.  The phosphate group it is also present in ambient bacterial DNA and RNA which measure more than one μm in diameter, and containing different chemical components in fraction term, thus giving us the ability to distinguish between the two weighting the different spectrum and flux contribution.

Most of the Phosphorus is present in the genetic strand of RNA and in particular in the phosphate group but trace amounts are also in viral proteins that contain the amino acid methionine. Phosphorus, on the other hand, is absent in the EPA pollutants listed in Table 1.

| PM and VOC* | | | |
|---|---|---|---|
| NOX | | | |
| SOX | | | |
| CO | | | |
| O3 | | | |
| **\*PM and VOC** | | | |
| 1,1,1-Trichloroethane | Bromornethane | 1,3-Butadiene | Farmaldehyde |
| 1,1,2,2-Tetrachloroethane | Carbon Disulfide | 2-Butanone | Gasoline, Automative |
| 1,1,2-Trichloroethane | Carbon Tetrachloride | 2-Hexanone | Hexachlorobutadiene |
| 1,1-Dichlorocthane | Chlorobenzene | Acetone | Hexachloroethane |
| 1,1-Dictioroethene | Chloroethane | Acrolein | Hydrazines |
| 1,2,3-Trichloropropane | chloroform | Benzene | Methyl Mercaptan |
| 1,2-Dibroma-3-Chloropropane | Chloromethane | Bromodichloromethane | n-Hexane |
| 1.2-Dibromaethane | Dichlorobenzenes | Scoddard Solvent | Nitrobenzene |
| 1,2-Dichloroethane | Dichloropropenes | Toluene | Styrene |
| 1,2-Dichloroethane | Ethylebenzene | Trichloroethylene (TCE) | Tetrachloroethylene |
| 1.2-Dichloropropane | Ethylene Dibromide | Vinyl Chloride | Xylenes |

**Table 1** PM and VOC Contamination

Therefore, Phosphorus, as part of the PO₄ group, has been assumed as a marker of no living matter, potentially a virus. A physical model has been designed in order to detect PO₄ groups amount changes related to different viral clusters. The present study aims to detect, by means of the particle Monte Carlo computer code Fluka 2020 [7-8] a cluster element spectrum with concentrations from 1 ppm in air to 0.1%, according to a composition of air in Table 1 and a simplified chemical form of a virus as a biological matter in air as reported in Figure 1 [1] with its own atom modelling.

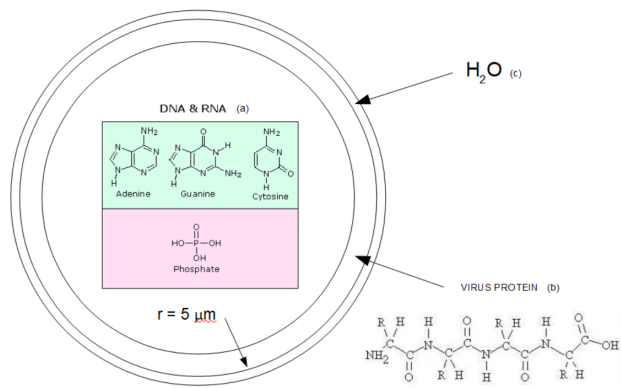

**Figure 1** Virion Particle Modelling

The proposed methodology is based on a subatomic coupled particle analysis (photon-electron) and the subsequent detection on top of their spectra (i.e., fluxes as a function of PO₄ group contamination ppm). The present study refers to previous works by the same authors [1,2] regarding RNA-DNA modelling and their identification through the interactions of particles such as muons and photons.

The purpose of this work is among some preliminary activities in view of experimental work to be performed using a photon beam with energy between 110-150 keV energy range due to a cross sections evaluation and its consequent discrimination ability on top of shielding requirements.

In order to justify the involved physics and the results in the 130 keV beam energy configuration a cross sections analysis for nitrogen and phosphorus content in the $PO_4$ group has been performed taking into account the EPDL97 Library [8]. Those two elements have a primary role from a photo atomic contribution point of view in the microscopic and macroscopic cross sections acting in the cluster mixture. The model proposed is based on multiple spherical geometry clusters in grid cells whose spheres (11) have different sizes (as a function of ppm viral contamination). However, those clusters are immersed in a 75% Nitrogen composition system which has a weighting factor as far as macroscopic cross section evaluation.

The microscopic cross section represents the effective target area of a single target nucleus for an incident particle and units are given in barn (1. barn = $10^{-24}$ cm$^2$) or cm$^2$, while the macroscopic cross-section represents the effective target area of all of the nuclei contained in the volume of the material and units are given in cm$^{-1}$. The macroscopic cross section can be obtained from the microscopic according to the following equation where N stands for nuclei density:

$$\Sigma = \sigma \cdot N$$

In a mixture case with different chemical elements, it is necessary to determine the macroscopic cross section for each isotope and therefore sum all the individual macroscopic cross-sections. Moreover, both factors (different atomic densities and different cross-sections) must be considered in the calculation of the macroscopic cross-section of the mixture taken into account.

The Avogadro's number $N_0 = 6.022 \times 10^{23}$, is the number of particles that is contained in the amount of substance given by one mole. Thus, if M is the molecular weight, the ratio $N_0/M$ equals to the number of molecules in 1g of the mixture. The number of molecules per cm$^3$ in the material of density $\varrho$ and the macroscopic cross-section for mixtures are given by following equations:

$$N_i = \varrho_i \cdot N_0 / M_i$$

$$\sum\nolimits_{mixt} = \sum\nolimits_I \times Ő_i$$

In the photon energy ($E_P$) interval 0.01 keV < $E_P$ < 1 MeV the main contribution to the total photon cross section for the Nitrogen is due to Rayleigh, Compton and Photoelectric effects. Moreover, in that energy interval the total photon cross section has a "higher trend" in terms of photon incident probability (from 0.1 to 5E+5 cm$^2$/g) due to the combination of the previous three effects with the dominant photoelectric component.

The probability involved is about a photon beam interacting with nuclei in the mixture in different layers such as:

$$-dI(x)/I(x) = \Sigma_t \cdot dx$$

Where $\Sigma_t$ stands for total cross section and $dI(x)$ is the number of photons interacting in dx, while $-dI(x)/I(x)$ is its probability to interact in the next layer.

From the previous equation follows the photon probability of interaction in dx such as:

$$P(x)dx = \Sigma_t dx \cdot e^{-\Sigma_t \cdot x} = \Sigma_t \, e^{-\Sigma_t \cdot x} \, dx$$

At 80 keV the Photoelectric and the Rayleigh effect drop leaving the Compton effect as dominant component with a cross section in the interval 0.0001 - 0.01 barn and decreasing until the pair production becomes the dominant component in the total photon cross section flatting the curve to 0.005 barn.

The photon beam source can be provided from an extraction line of a linear/circular electron accelerator. In this case, a simple bremsstrahlung target interaction with the electron beam produces the requested photons. Another possibility is to obtain the photons of fixed energy from a selected radioactive source by natural radioisotope decay.

## 2. Materials and Methods

The physical model under analysis and its simulation by Fluka 2020 particles computer code is based on a photon beam source of 130 keV accessible from an extraction line of an electron accelerator. The beam interacts with a cylindrical sample volume—with the axis on x—of air of radius r = 5 cm and height h = 10 cm as *s* sample tank (Figure 2 to 4), which is analyzed at x = 10 cm through a double plates' ionization chamber detector or with a fully depleted pnCCD, a special type of charge coupled devices detector with a CsI(Tl) scintillator .

The sample air chemical composition considered for the present and previous studies is shown in Table 2 [1, 9].

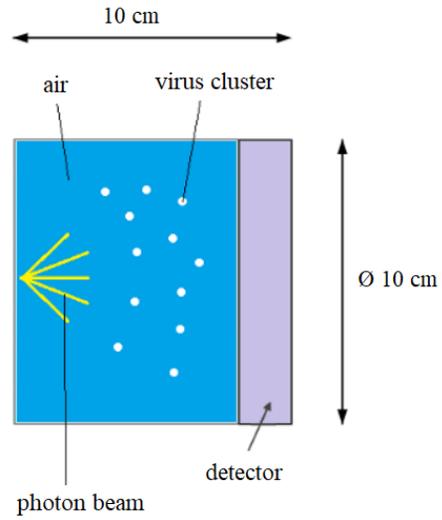

**Figure 2** Physical model x-z section of air sample and Virion Particles

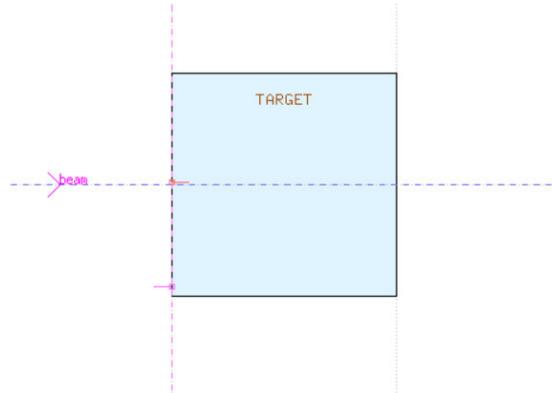

**Figure 3** Geometry Model

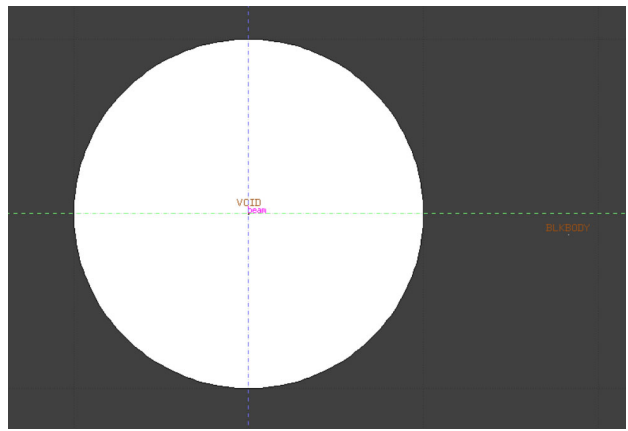

**Figure 4** Geometry Region Modelling -Target, Void, Black hole

|  | 0 ppm |
|---|---|
| *Oxygen* | 0.2317812 |
| *Nitrogen* | 0.7552670 |
| *Argon* | 0.0128270 |
| *Carbon* | 0.0001248 |
|  |  |
| Norm | 1.0000000 |

Table 2 Air Composition Sample Tank (0 ppm contamination)

Among all possible subatomic secondary particles generated, only photons (coming from 1st and 2nd fluorescence, bremsstrahlung) and electrons were considered (coming from Compton recoil, photo electric, photon Auger, electron Auger and knock-on), as other particles are negligible.

The photoelectric effect, which is the predominant reaction in the electron creation, consists of the absorption of the incident photon, with energy $E$, with emission of several fluorescent photons and the ejection or excitation of an orbital electron of binding energy $e < E$. Photons of first fluorescence are emitted with energy greater than 1 keV; those ones of second fluorescence are still greater than 1 keV and are caused by residual excitation of the first fluorescence process, leading to a second emission.

The analysis considered both electrons and photons, as said before, without neglecting any secondary photon production by performing a photon/electron coupled calculation and by keeping track of the electron/photon mean free path in the media. All the results proposed concern the photon fluxes and spectra (i.e., fluxes as a function of photon energy) of interest, where also all the possible primary and secondary electrons into the sample volume were taken into account.

The virion particles have dimension 20-250 nm and have been assumed incapsulated in lipidic coat as well as in a water film in order to modelling an "expectoration human process". The virions are described in 11 cluster configurations through a volumetric cell grid (Figures 6 to 9); each cluster is composed of microspheres with a radius of 5 µm and a volume of $5.24 \times 10^{-7}$ mm$^3$ per incapsulated virion particle, with a mutual distance of $1 < d_i < 9$ cm among the clusters along all the axes and evaluated in the air sample tank at different concentrations from 1 ppm up to 1,000 ppm (Table 3 to 8); this has been taken into account in the physical modelling with multiple layers radius: $\sum (a_i, b_i, c_i)$ as a function of the number of particles composing each cluster as index i=1-1,000 ppm (Figure 6 to 9). In the geometrical model the PO$_4$ group analysis outcome has been analyzed by control check volumes/surfaces in order to evaluate energy distributions and particle mean free path (photon and electron) (yellow squares, Figure 6)

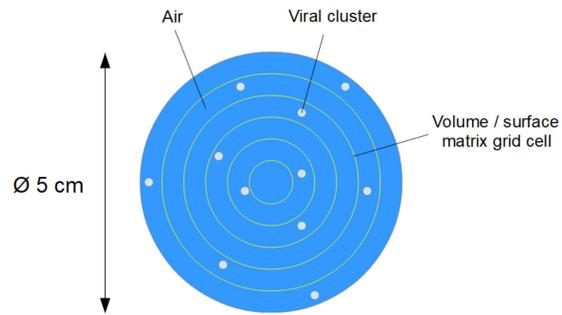

**Figure 5** Geometrical model x-z section

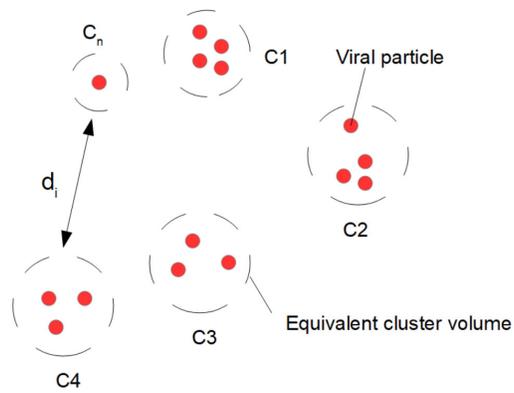

**Figure 6** Cluster Configuration

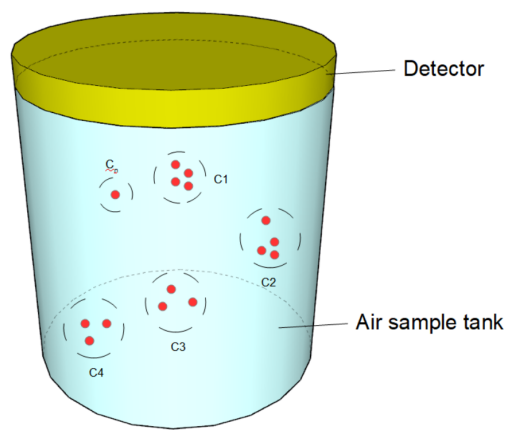

**Figure 7** Volumetric Cluster Cells in 3D - "The Babi Mini Mug"

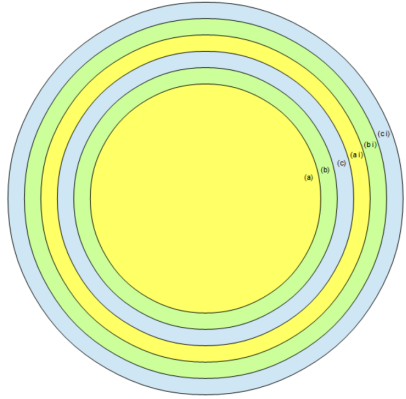

**Figure 8** Virion Cluster, x-z section model

|  | 0 ppm | 1 ppm | 10 ppm | 100 ppm | 1000 ppm |
|---|---|---|---|---|---|
| *Oxygen* | 0.2317812 | 0.2317810 | 0.2317789 | 0.2317580 | 0.2315494 |
| *Nitrogen* | 0.7552670 | 0.7552662 | 0.7552594 | 0.7551915 | 0.7545117 |
| *Argon* | 0.0128270 | 0.0128270 | 0.0128269 | 0.0128257 | 0.0128142 |
| *Carbon* | 0.0001248 | 0.0001248 | 0.0001248 | 0.0001248 | 0.0001247 |
| *PO4 Group* | 0.0000000 | 0.0000010 | 0.0000100 | 0.0001000 | 0.0010000 |
| Norm | 1.0000000 | 1.0000000 | 1.0000000 | 1.0000000 | 1.0000000 |

**Table 3** Air Vs PO4 Virion ppm

|  | (1 ppm) | (10 ppm) | (100 ppm) | (1000 ppm) |
|---|---|---|---|---|
| **Cluster N** | *ppm per cluster* | *ppm per cluster* | *ppm per cluster* | *ppm per cluster* |
| 1 | 0.1 | 1 | 10 | 100 |
| 2 | 0.05 | 0.5 | 5 | 50 |
| 3 | 0.2 | 2 | 20 | 200 |
| 4 | 0.13 | 1.3 | 13 | 130 |
| 5 | 0.19 | 1.9 | 19 | 190 |
| 6 | 0.03 | 0.3 | 3 | 30 |
| 7 | 0.08 | 0.8 | 8 | 80 |
| 8 | 0.04 | 0.4 | 4 | 40 |
| 9 | 0.02 | 0.2 | 2 | 20 |
| 10 | 0.09 | 0.9 | 9 | 90 |
| 11 | 0.07 | 0.7 | 7 | 70 |
| Norm | 1 | 10 | 100 | 1000 |

**Table 4** Cluster Configuration Vs ppm

| Cluster N | (1 ppm) ppm per cluster | (1 ppm) % ppm cluster | Particles N per cluster | Volume (mm3) per cluster | Eq. radius (mm) |
|---|---|---|---|---|---|
| 1 | 0.1 | 10.00% | 1.048E+05 | 5.493E-02 | 0.23581 |
| 2 | 0.05 | 5.00% | 5.241E+04 | 2.746E-02 | 0.18716 |
| 3 | 0.2 | 20.00% | 2.096E+05 | 1.099E-01 | 0.29710 |
| 4 | 0.13 | 13.00% | 1.363E+05 | 7.140E-02 | 0.25736 |
| 5 | 0.19 | 19.00% | 1.992E+05 | 1.044E-01 | 0.29207 |
| 6 | 0.03 | 3.00% | 3.145E+04 | 1.648E-02 | 0.15786 |
| 7 | 0.08 | 8.00% | 8.386E+04 | 4.394E-02 | 0.21891 |
| 8 | 0.04 | 4.00% | 4.193E+04 | 2.197E-02 | 0.17375 |
| 9 | 0.02 | 2.00% | 2.096E+04 | 1.099E-02 | 0.13790 |
| 10 | 0.09 | 9.00% | 9.434E+04 | 4.943E-02 | 0.22767 |
| 11 | 0.07 | 7.00% | 7.338E+04 | 3.845E-02 | 0.20938 |
| Norm | 1 | 100.00% | 1.048E+06 | 5.493E-01 | 2.39498 |

**Table 5** Cluster Configuration - 1 ppm

| Cluster N | (10 ppm) ppm per cluster | (10 ppm) % ppm cluster | Particles N per cluster | Volume (mm3) per cluster | Eq. radius (mm) |
|---|---|---|---|---|---|
| 1 | 1 | 10.00% | 1.048E+06 | 5.493E-01 | 0.50804 |
| 2 | 0.5 | 5.00% | 5.241E+05 | 2.746E-01 | 0.40323 |
| 3 | 2 | 20.00% | 2.096E+06 | 1.099E+00 | 0.64009 |
| 4 | 1.3 | 13.00% | 1.363E+06 | 7.140E-01 | 0.55447 |
| 5 | 1.9 | 19.00% | 1.992E+06 | 1.044E+00 | 0.62924 |
| 6 | 0.3 | 3.00% | 3.145E+05 | 1.648E-01 | 0.34010 |
| 7 | 0.8 | 8.00% | 8.386E+05 | 4.394E-01 | 0.47162 |
| 8 | 0.4 | 4.00% | 4.193E+05 | 2.197E-01 | 0.37433 |
| 9 | 0.2 | 2.00% | 2.096E+05 | 1.099E-01 | 0.29710 |
| 10 | 0.9 | 9.00% | 9.434E+05 | 4.943E-01 | 0.49051 |
| 11 | 0.7 | 7.00% | 7.338E+05 | 3.845E-01 | 0.45109 |
| Norm | 10 | 100.00% | 1.048E+07 | 5.493E+00 | 5.15982 |

**Table 6** Cluster Configuration - 10 ppm

| Cluster N | (100 ppm) ppm per cluster | (100 ppm) % ppm cluster | Particles N per cluster | Volume (mm3) per cluster | Eq. radius (mm) |
|---|---|---|---|---|---|
| 1 | 10 | 10.00% | 1.048E+07 | 5.493E+00 | 1.09454 |
| 2 | 5 | 5.00% | 5.241E+06 | 2.746E+00 | 0.86874 |
| 3 | 20 | 20.00% | 2.096E+07 | 1.099E+01 | 1.37903 |
| 4 | 13 | 13.00% | 1.363E+07 | 7.140E+00 | 1.19457 |
| 5 | 19 | 19.00% | 1.992E+07 | 1.044E+01 | 1.35566 |
| 6 | 3 | 3.00% | 3.145E+06 | 1.648E+00 | 0.73272 |
| 7 | 8 | 8.00% | 8.386E+06 | 4.394E+00 | 1.01608 |
| 8 | 4 | 4.00% | 4.193E+06 | 2.197E+00 | 0.80646 |
| 9 | 2 | 2.00% | 2.096E+06 | 1.099E+00 | 0.64009 |
| 10 | 9 | 9.00% | 9.434E+06 | 4.943E+00 | 1.05677 |
| 11 | 7 | 7.00% | 7.338E+06 | 3.845E+00 | 0.97185 |
| Norm | 100 | 100.00% | 1.048E+08 | 5.493E+01 | 11.11651 |

Table 7 Cluster Configuration - 100 ppm

| Cluster N | (1000 ppm) ppm per cluster | (1000 ppm) % ppm cluster | Particles N per cluster | Volume (mm3) per cluster | Eq. radius (mm) |
|---|---|---|---|---|---|
| 1 | 100 | 10.00% | 1.048E+08 | 5.493E+01 | 2.35811 |
| 2 | 50 | 5.00% | 5.241E+07 | 2.746E+01 | 1.87164 |
| 3 | 200 | 20.00% | 2.096E+08 | 1.099E+02 | 2.97104 |
| 4 | 130 | 13.00% | 1.363E+08 | 7.140E+01 | 2.57363 |
| 5 | 190 | 19.00% | 1.992E+08 | 1.044E+02 | 2.92067 |
| 6 | 30 | 3.00% | 3.145E+07 | 1.648E+01 | 1.57860 |
| 7 | 80 | 8.00% | 8.386E+07 | 4.394E+01 | 2.18908 |
| 8 | 40 | 4.00% | 4.193E+07 | 2.197E+01 | 1.73747 |
| 9 | 20 | 2.00% | 2.096E+07 | 1.099E+01 | 1.37903 |
| 10 | 90 | 9.00% | 9.434E+07 | 4.943E+01 | 2.27673 |
| 11 | 70 | 7.00% | 7.338E+07 | 3.845E+01 | 2.09378 |
| Norm | 1000 | 100.00% | 1.048E+09 | 5.493E+02 | 23.94978 |

Table 8 Cluster Configuration - 1000 ppm

The simulations were performed step by step in different cluster stages: Stage 1, with 0 ppm contamination to investigate the physics involved in the basic case without viral particles; Stage 2, evaluating an escalating contamination grade as maximum stress test of 1 ppm, 10 ppm, 100 ppm, 1000 ppm to determine the subatomic particles' stopping power and the shielding effects that give the photon/electron fluences and energy spectra in different cluster contaminations with an average relative error of 3% counted on seven independent runs. Each run has been tuned, in term of number histories, based on the deviation of the average relative error which can be detected by error spikes in few energy bins due to lack of sufficient particle histories. The tuning process just mentioned had the capability to converge our statistic and to assure seven independent runs are adequate. Although also the 1000 ppm case was simulated, we didn't report it in the main results because a case of non-interest inasmuch as interested in low contamination detection less than 10 ppm.

**3. Results**

Here we present the results of the analysis showing the photon/electron fluences and energy spectra of the Monte Carlo simulations in the presence of virion cluster contaminations and without it in the detector chamber, located at $x$ = 10 cm on the top of the sample tank on the x-axis, as shown in Figure 2-8.

The study analyzed the photon/electron fluences and their contributions at different virion cluster grades with different energy spectrum peaks due to cross-section considerations and energy spectrum degradation

Every photon spectrum has been subdivided in 3 energy bands: Low, Medium, High in order to highlight the differences in the channel acquisition as a function of the contamination.

As far as the photon spectra as a function of viral contamination in the air sample, the channels:

- 2 keV, 3 keV, 4 keV, 20 keV, 90 keV are markers as a function of the $PO_4$ Group ppm contamination

Concerning electron spectra as a function of viral contamination in the air sample, the channels:
- 7 keV, 8 keV, 10 keV, 20 keV are markers as a function of the $PO_4$ Group ppm contamination

Figure 10 shows the photon/electron fluences as a function of viral contamination coming from the $PO_4$ group ppm contamination.

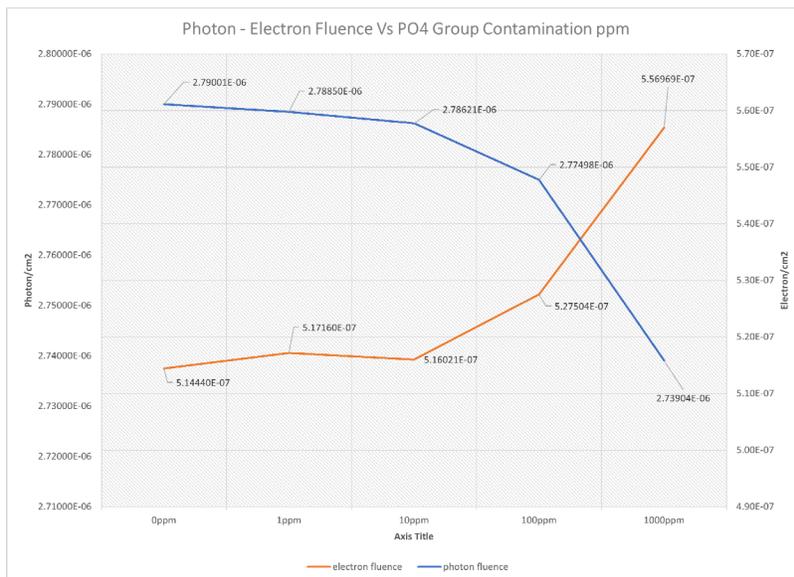

**Figure 9** Fluences - Photon and Electron Vs Contamination

As mentioned in Section 3, the graphs in Figures 11 to 19 show the photon/electron energy spectra and their particle counts as a function of viral particles in cluster configuration.

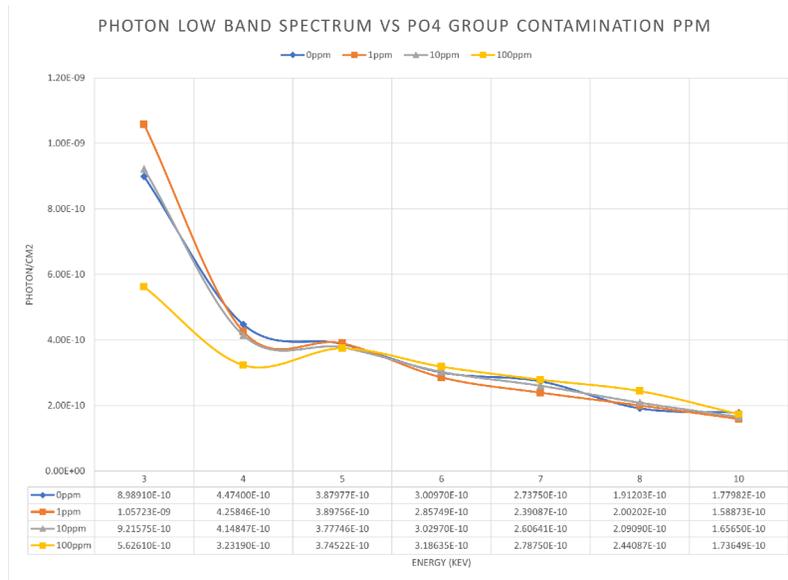

**Figure 10** Low Band Photon Spectrum A - Air Sample Vs Contamination

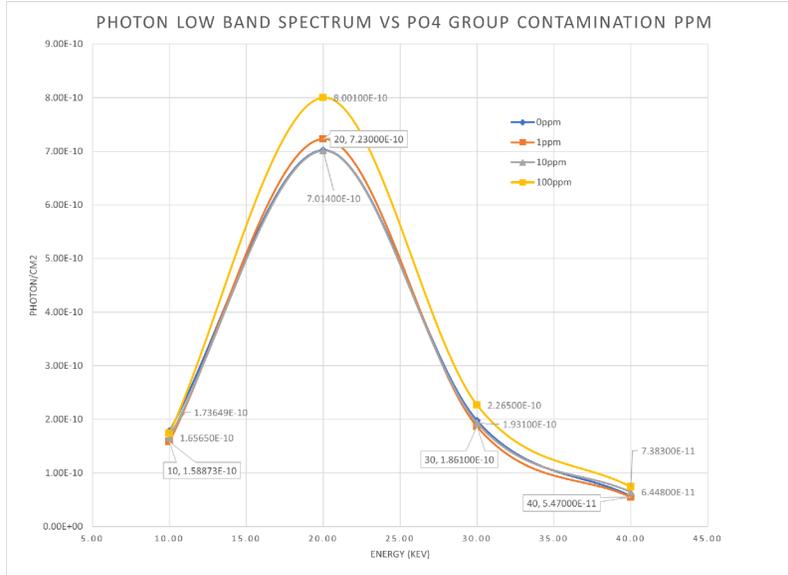

**Figure 11** Low Band Photon Spectrum B - Air Sample Vs Contamination

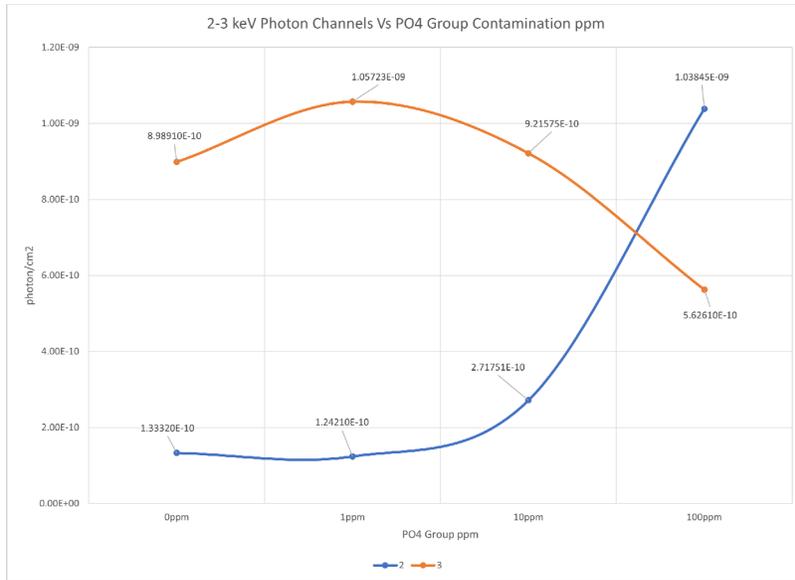

**Figure 12** Photon Channel 2-3 keV Vs Contamination

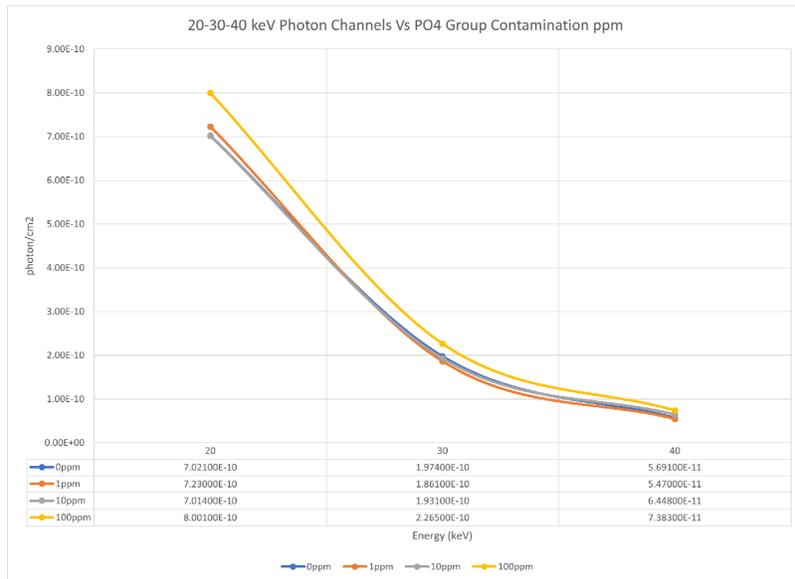

**Figure 13** Medium Band Photon Spectrum - Air Sample Vs Contamination

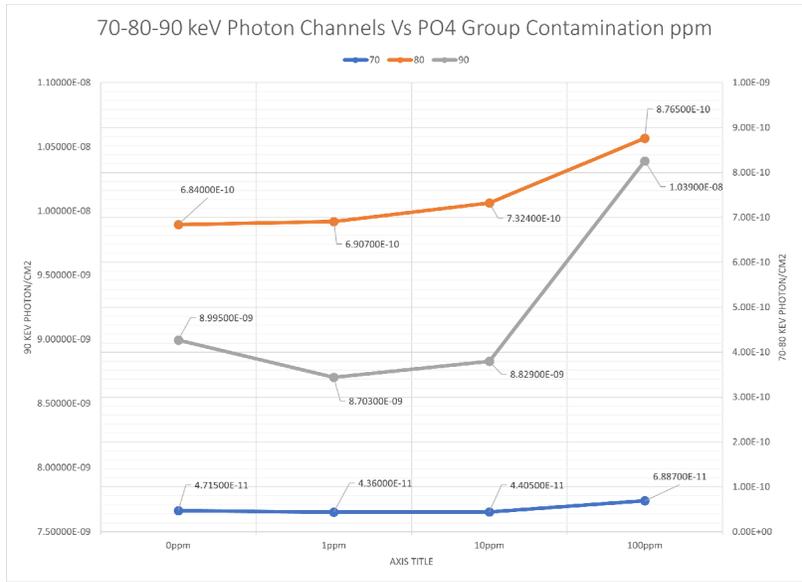

**Figure 14** High Band Photon Spectrum B1 - Air Sample Vs Contamination

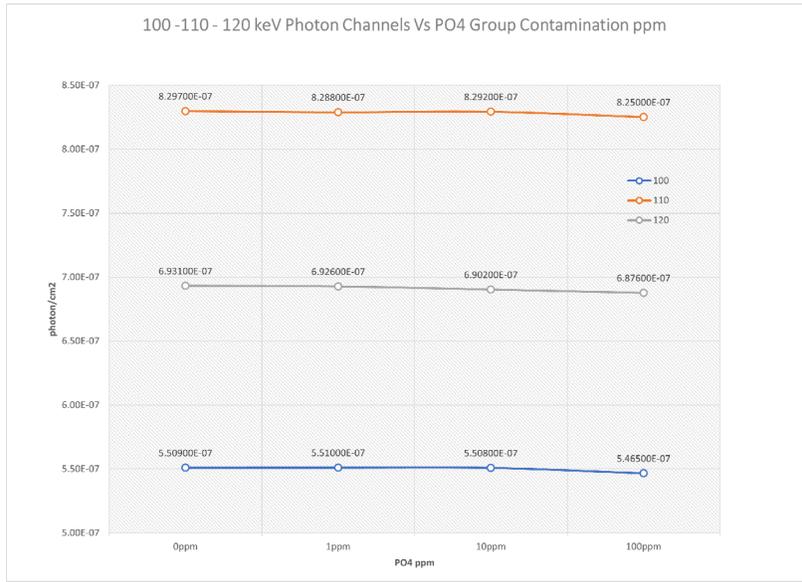

**Figure 15** High Band Photon Spectrum B2 - Air Sample Vs Contamination

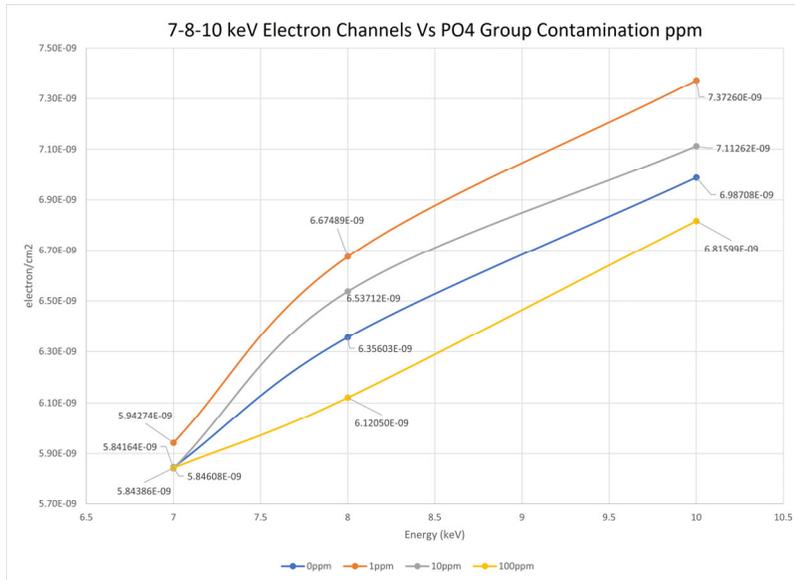

**Figure 16** Low Band Electron Spectrum - Air Sample Vs Contamination

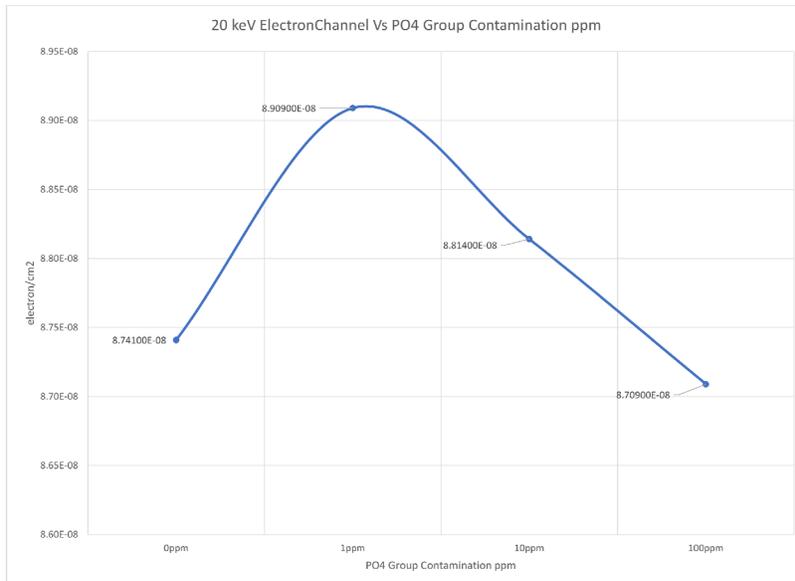

**Figure 17** Electron 20 keV Channel - Air Sample Vs Contamination

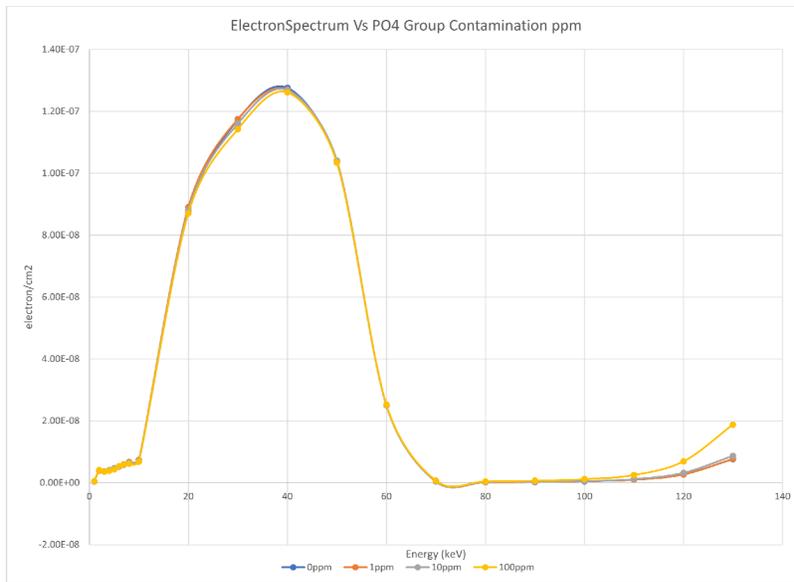

**Figure 18** Electron Spectrum - Air Sample Vs Contamination

## 4. Discussion

Given the current pandemic as a result of SaRSCoV-2, causing CoVID-19, and the possibility of more viruses crossing species boundaries resulting in yet more viral epidemics, the need for detection of these infectious agents is critical to controlling current and future outbreaks of infectious diseases caused by viruses. Here we have reported that the technology for detecting these infectious agents could exist and could result in quantifiable measurements of ambient viral particles [2].

Using the physics described in this paper, it could be possible to detect specific viral particles in the environment. This can provide crucial, necessary for the healthcare community to detect viral levels and correlate those to the probability of individuals in a defined area to become infected. Detection of particles in ambient air is crucial in order to lower infection rates in places such as hospitals, retirement homes, and other healthcare facilities. Furthermore, this technology would enable business and government official to make informed decisions about the appropriate measures necessary to take in order to lower infection rates.

As climate change continues and arthropod (vectors) populations migrate, this technology could enable the tracking and identification of the movement of infectious viruses to different geographical regions populated by humans. "According to a report by the U.S. Centers for Disease Control and Prevention. It found vector-borne diseases spread by parasitic insects and arachnids more than tripled in the U.S. over 12 years — from 27,388 cases in 2004 to 96,075 in 2016" [10].

The photon/electron fluences and spectra can discriminate the amount of virion particles contamination by using its own "particle signature" in terms of photon/electron counts at the detector point combined with the spectrum analysis, as reported. This will enable humans to track the possibility of future viral outbreaks.

The model proposed is based on multiple spherical geometry clusters in grid cells whose spheres have different sizes (as a function of ppm viral contamination). However, those clusters are immersed in a 75% Nitrogen composition system which has weighting factor as far as macroscopic cross section contribution.

On Figures 11-12-17 at low contamination 1-100 ppm the photon/electron count profiles show self-shielding and stopping powers behaviors as a function volumetric content as shown in Figure 19.

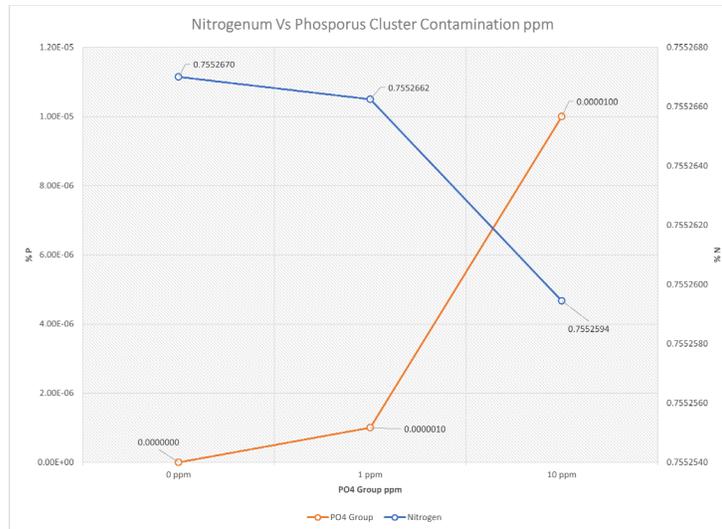

**Figure 19** N Vs P Cluster Contamination 0-10 ppm

As shown in Figures from 17 to 19, the electron fluences and spectra associated with the sample of air at different concentrations of the viral particles show a trend in term of electron/cm$^2$ on different detection energy channels: 7-8-10-20 keV are identifiers as function of PO$_4$ ppm contamination as much as 120-130 keV. In all the other electron channels superimposed conditions are present, from an energy spectrum point of view, and not enough to discriminate and evaluate their own contribution as function of ppm contamination.

As shown in the cross-section graphs (Figures 20-21), a 130 keV photon beam allows the identification of Phosphorus, as part of the PO$_4$ Group in the cluster configurations.

Analyzing all the significative channels as "particle signature" found in the Fluka simulations it is possible to confirm that: at 2-3 keV channels, Nitrogen has a 12 barn cross-section while Phosphorus has a cross-section of about 54 barn ; therefore, at low energy between 1-10 keV the identification of a virus (that contains Phosphorus as a PO$_4$ group) counterbalances a low concentration in ppm as deficit (in proportion) of the first and second fluorescence. Those results are also confirmed by the electron spectrum on the same channels due to photoelectric effect, 1$^{st,}$ and 2$^{nd}$ fluorescence.

By increasing the energy channels to 20 keV, Nitrogen has 0.4 barn while Phosphorus has a value of 4 barn. As far as 90 keV Nitrogen has a value of 0.002 barn while Phosphorus 0.03 barn.

To summarize, for primary photon beam energies less than 1 keV the Nitrogen total photon cross section has an overall higher probability (in barn) compared to the Phosphorus in the same energy range, leading to difficulties in the discrimination process. For energies greater than 1 MeV the steady decreasing of the Compton effect and increasing of Pair production make a potential analysis subtle and almost imperceptible. To avoid this problem, it is necessary to proceed with a photon beam between 130 and 150 keV at an experimental level in order to have the ability to discriminate the PO$_4$ group since the dominant volumetric components would be Nitrogen and Oxygen. In Figure 22 a potential new application design and set-up is also shown.

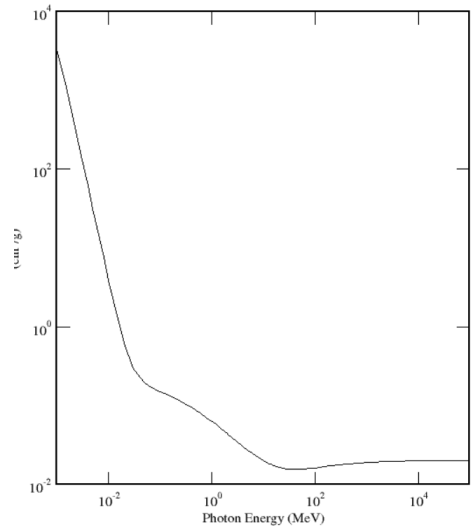

**Figure 20** Nitrogen Photon Cross Section

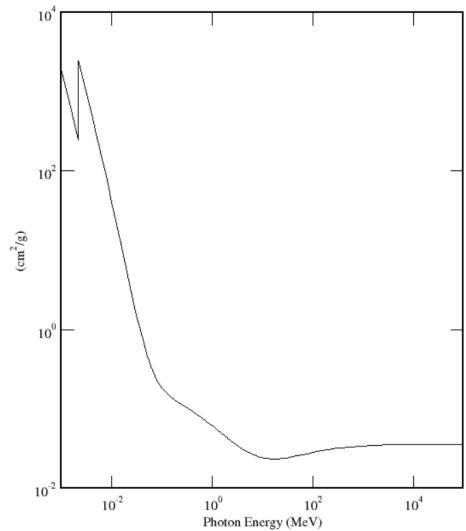

**Figure 21** Phosphorus Photon Cross Section

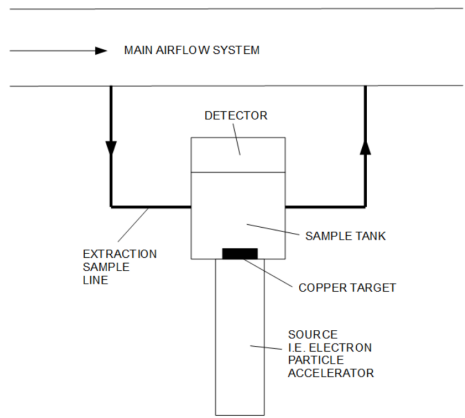

**Figure 22** Detection - Design and set-up

## 5. Conclusions

This study proposes a new approach to identify low contaminations of viral particles clusters mixed in air using Monte Carlo simulations. The study evaluates the primary/secondary photon/electron energy spectra and fluxes revealed by a potential experimental detector on different energy channels.

Different types of contamination grades can be discriminated using their trends Vs photon/s*cm$^2$ - electron/s*cm$^2$ evaluated on different energy intervals as a function of the energy photon beam primary source. Every single contamination is unique in its own spectrum photon/electron signature, and the flux acts as a unique identifier in the detection process so that it can give the ppm amount of virion particles in the air.

Finally, this work has shown that there is still a need for experimental results to confirm the computational analysis to identify RNA in air. Experimental, theoretical, and computational work are all needed for a better understanding of the link between the physical nature of a radiation and its effects for the RNA identification and detection as aerosol in air samples by means of photon/electron interactions.

**Patent N/A**

**Author Contributions:** Conceptualization, L.J.T., P.N.; methodology, L.J.T, P.N.; software, L.J.T, P.N.; validation, L.J.T., J.I.A.; investigation, R.S, J.I.A..; data curation, R.S.; writing—original draft preparation, P.N., L.J.T.; writing—review and editing, L.J.T, P.N., J.I.A.; visualization, R.S.; All authors have read and agreed to the published version of the manuscript.

**Funding:** This research received no external funding.

**Acknowledgments:** We deeply thank Giulio Magrin, Ilaria A. Valli, Alessandro Alemberti.

**Conflicts of Interest:** The authors declare no conflict of interest.

*In memory of Alberto Negrini, Augusto Solei and Mario Tagliapietra*